\documentclass[12pt]{article}
\newcommand{\la}[1]{\label{#1}}

\newcommand{\vecn}{{\vec{\bf n}}}
\newcommand{\vecm}{{\vec{\bf m}}}
\newcommand{\vecu}{{\vec{\bf u}}}
\newcommand{\vecc}{{\vec{\bf c}}}
\newcommand{\vect}{{\vec{\bf t}}}

\newcommand{\vece}{{\vec{\bf e}}}
\newcommand{\vecx}{{\vec{\bf x}}}
\newcommand{\veccx}{{\vec{\bf X}}}
\newcommand{\vecb}{{\vec{\bf b}}}

\newcommand{\matr}{{\bf R}}
\newcommand{\mats}{{\bf S}}

\newcommand{\be}{\begin{equation}}
\newcommand{\ee}{\end{equation}}
\newcommand{\ba}{\begin{eqnarray}}
\newcommand{\ea}{\end{eqnarray}}
\newcommand{\bastar}{\begin{eqnarray*}}
\newcommand{\eastar}{\end{eqnarray*}}

\begin{document}
\begin{titlepage}

\vskip 2.0truecm

\begin{center}
$ ~$
\end{center}


\begin{center}
{ 
\bf \large \bf FIELD THEORIES FROM BUNDLES OF STRINGS \\ 
}
\end{center}

\vskip 1.5cm

\begin{center}
{
\bf Sazzad Nasir$^{\, \star}$  \ 
and \  \bf Antti J. Niemi$^{\, \sharp }$ 
} \\

\vskip 0.6cm

{\it Department of Theoretical Physics,
Uppsala University \\
P.O. Box 803, S-75108, Uppsala, Sweden } \\

\end{center}

\vskip 2.5cm
\noindent
We propose that at low energy four dimensional 
bosonic strings may form bound states where they 
become bundled together much like the filaments 
in a cable. We inspect the properties of these bundles
in terms of their extrinsic geometry. This involves
both torsion and curvature contributions, and leads to four 
dimensional field theories where the bundles appear 
as closed knotted solitons. Field theories that allow 
for this interpretation include the Faddeev model, the 
Skyrme model, the Yang-Mills theory, and a Hartree-type field 
theory model of a two-component plasma. 
\vfill

\begin{flushleft}
\rule{5.1 in}{.007 in} \\
$^{\star}$  {\small E-mail: \scriptsize
\bf NASIR@TEORFYS.UU.SE} \\
\hskip 0.5cm supported by the G\"oran Gustafsson Stiftelse \\
$^{\sharp}$  {\small E-mail: \scriptsize
\bf NIEMI@TEORFYS.UU.SE} \\
\hskip 0.5cm supported by NFR Grant F-AA/FU 06821-308 \\

\end{flushleft}
 
\end{titlepage}

A closed bosonic string ${\cal S}$ describes 
the motion of a circle $\mats^1$ in a D-dimensional 
manifold. The dynamics of ${\cal S}$ is 
determined by the Polyakov action which 
governs the behavior of strings at 
very high energies. But at lower energies there 
can be corrections, which are due to terms
that characterize
the extrinsic geometry of the string 
\cite{poly}. In three spatial 
dimensions this involves
the curvature $\kappa$ and the 
torsion $\tau$ of ${\cal S}$ when the string is
viewed as a curve which is embedded in $\matr^3$.
For this we consider a string described  
by coordinates $\vecx(s)\in \matr^3$ that are  
parametrizes by the arc-length $s \in [0,L]$ 
of the curve. The total
length $L$ can be variable, and since the 
string is closed $\vecx(s+L) = \vecx(s)$. The 
unit tangent to the curve $\vecx(s)$ 
\be
\vect \ = \ \frac{d \vecx}{ds}
\la{tang}
\ee
then leads to the Frenet equations
\[
d\vect \ = \ \frac{1}{2}\kappa (\vecc_+ + \vecc_-) ds
\]
\be
d\vecc_\pm \ = \ -(\kappa \vect \pm i \tau \vecc_\pm )ds
\la{fre}
\ee
where $\vecc_{\pm} = \vecn \pm i \vecb$ and 
$\vecn$ is the unit normal vector and $\vecb$
is the unit binormal vector. The curvature 
$\kappa$ and torsion 
$\tau$ are
\be
\kappa \ = \ \frac{1}{2} 
\vecc_{\pm} \cdot \frac{d\vect}{ds}
\la{kappa}
\ee
\be
\tau \ = \ \frac{i}{2} \vecc_- \cdot \frac{d\vecc_+}{ds}
\la{tau}
\ee 
In the classical theory of elastic rods one
employs (\ref{kappa}) and (\ref{tau}) to 
define an energy for an individual
string $\vecx(s)$. Often one selects the
energy to contain at least the following 
terms
\be
E \ = \ c_L L \ + \ 
\int\limits_0^L ds ( c_\kappa \kappa^2 \ + \
c_\tau \tau^2 )
\la{kirc}
\ee
Here the first term relates to the Polyakov action
in the static case, and 
the $c_{L,\kappa,\tau}$ are some parameters.

In the present Letter we are interested in the
collective properties of such closed strings ${\cal S}$
in $\matr^3$. We are 
particularly interested in the possible existence of
configurations where the 
strings become smoothly bundled together, much 
like individual filaments are bundled together to
form a cable. Specifically,
we wish to generalize (\ref{kirc}) to an energy functional
that describes such an entire bundle of strings
in terms of natural geometrical quantities like
the concepts of torsion and curvature. Hypothetically,
such a bundle of strings could become 
relevant {\it e.g.} in describing an
effective low-energy theory of strings, at energy
scales where individual strings condense into bundled
bound states. We shall argue that this picture
leads to a number of familiar field theory models.

{\it A priori} there can be several distinct 
alternatives to bundle the strings together
and the ensuing dynamics certainly 
depends on the approach selected.
Here we shall employ a smooth three-component 
unit vector field $\vecm(\vecx) $ which maps $\matr^3$ 
to an internal two-sphere $\mats^2_{\bf m}$. 
This will lead to
a field theory description with a 
relativistic dynamics, in terms
of geometrical quantities that relate 
to the properties of the individual strings
in a natural fashion.
We are interested in configurations
that are localized in $\matr^3$ so that at large 
distances $\vecm$ approaches a
constant vector. Without any loss
of generality we select the asymptotic vector 
to point along the negative $z$-axis
$\vecm(\vecx) \to - \vece_3$.
The region where $\vecm$ deviates from 
the asymptotic value 
$- \vece_3$ then describes our bundle of 
strings. Such a localized $\vecm$ 
is a mapping from the compactified 
three-space $\matr^3 \sim \mats^3$ 
to the internal $\mats_{\bf 
m}^2$ so that the pre-image of any point in 
$\mats^2_{\bf m}$ corresponds to some closed string 
in $\matr^3$. These are the curves in $\matr^3$
along which $\vecm(\vecx)$ is constant
\be
\frac{d\vecm}{ds} \ = \
\frac{d x^i}{ds} \frac{\partial \vecm}{\partial x^i}
\ \equiv \ t_i  \partial_i \vecm \ = \ 0
\la{inde1}
\ee
with $t_i$ the unit tangent vector to the curve 
$\vecx(s)$.  

In the general 
case the individual strings ${\cal S}$ can be linked,
so that the bundle forms a knotted configuration in
$\matr^3$ which we assume to be smooth. 
Its self-linking number
is computed by the Hopf invariant of $\vecm$
which coincides with the Gauss linking number for any 
pair of the closed strings ${\cal S}$.
To compute the Hopf invariant we need to
describe the curves along which $\vecm$ 
remains constant in $\matr^3$.  For this we consider
a four component unit vector
$\psi_\mu(\vecx): \ \matr^3 \sim \mats^3
\to \mats^3$ that embeds the bundle in 
the configuration space $\matr^3 
\sim \mats^3$, and introduce the combinations
\[
z_1(\vecx) \ = \ \psi_1 (\vecx) + i \psi_2 (\vecx)
\]
\[
z_2 (\vecx) \ = \ \psi_3(\vecx) + i \psi_4(\vecx)
\]
We can select $\psi_\mu$ so that it 
yields the unit vector $\vecm$ by
\be
\vecm \ = \ Z^\dagger \vec {\bf \sigma } Z
\la{fibr}
\ee
where $Z \ = \ (z_1, z_2)$. Notice that this 
does not define $Z$ uniquely but there is a
U(1) gauge symmetry, the relation 
(\ref{fibr}) remains intact 
when we multiply $Z$ by a phase
\be
Z \ \to \ e^{\frac{i}{2}\xi} Z
\la{u1a}
\ee
When we parametrize
\be
Z \ = \  \pmatrix{ e^{i \phi_{12}} 
\sin \frac{1}{2}\vartheta \cr
e^{i \phi_{34}} \cos \frac{1}{2} \vartheta }
\la{para}
\ee
and set $\beta = \phi_{34} - \phi_{12}$ and $
\gamma = \pi - \vartheta$ we get
\be
\vecm \ = \  \pmatrix{ \cos \beta \sin \gamma
\cr \sin\beta \sin \gamma \cr \cos \gamma
}
\la{parm}
\ee
Hence the bundle resides in the region in 
$\matr^3$ where $\gamma \not= \pi$. 
Since $\vecm$ is independent of $\alpha = \phi_{12} +
\phi_{34}$ we identify $\alpha(\vecx)$ as a
coordinate generalization of the parameters $s$.
It is the bundle coordinate that describes 
the individual strings ${\cal S}$ 
in the directions $\vect$, the
curves in $\matr^3$ which are
traced by constant values of $\vecm(\vecx)$. 
The U(1) gauge transformation (\ref{u1a}) 
sends  $\alpha \to \alpha + \xi$ which relates to the
reparametrizations $s \to {\tilde s}(s)$ of 
the individual strings.
Since the physical properties of the bundle should 
be reparametrization invariant, any physical 
quantity should reflect an invariance 
under the gauge transformation (\ref{u1a}).

We define 
\be
A_i \ = \ i ( \partial_i Z^\dagger Z - 
Z^\dagger \partial_i Z)
\la{Ai}
\ee
We substitute (\ref{para}) in (\ref{Ai}) 
and combine the individual terms as follows,
\be
A \ = \ \cos \gamma d \beta \ + \
d \alpha
\la{Ai2}
\ee
The U(1) gauge transformation (\ref{u1a})
sends $A$ into
\be
A \ \to \ \cos \gamma d \beta \ + \
d ( \alpha + \xi)
\la{u1a2}
\ee
which identifies $A$ as the corresponding 
gauge field. Its exterior 
derivative yields the pull-back of
the area two-form on $\mats^2_{\bf m}$,
\be
F \ = \ dA \ = \
\sin \gamma d \beta \wedge d \gamma
\ = \ - \frac{1}{2}
\vecm \cdot d\vecm \wedge d \vecm
\la{Fij2}
\ee
and the dual one-form with components
\[
B_{i} \ = \ \frac{1}{2} \epsilon_{ijk} F_{jk}
\]
is parallel to the tangents of the 
individual strings $\vecx(s)$ since
\[
\epsilon_{ijk}t_j B_k  \ = \ 0
\]
The Hopf invariant is
\[
Q_{H} \ = \ \frac{1}{8\pi^2} \int F \wedge A \ = \ 
\frac{1}{8\pi^2} \int \sin \gamma \ d\alpha 
\wedge d \beta \wedge d \gamma
\]
If the Hopf invariant is nonvanishing the bundle
forms a nontrivial knot. In that case
the flat connection $d\alpha$ can not be 
entirely removed by the gauge 
transformation (\ref{u1a2}) since the Wilson loops 
\[
\oint_{\cal S} d\alpha \ = \ \int_0^L ds \, t_i 
\partial_i \alpha
\]
along the closed strings ${\cal S}$ in the bundle
are necessarily nontrivial.

In order to obtain appropriate generalizations of the
curvature and torsion (\ref{fre}) and in particular
the ensuing version of (\ref{kirc}) we consider a 
generic point $\vecm
\in \mats^2_{\bf m}$. 
Its pre-image corresponds
to a generic string ${\cal S}$ in the bundle. 
We introduce the
tangent vectors $\vece_{\pm} = 
\vece_1 \pm i \vece_2$ of $ \mats^2_{\bf m}$ in
$\matr^3_{\bf m}$
so that $(\vece_1, \vece_2, 
\vecm)$ is a right-handed orthonormal triplet. 
When we parametrize $\vecm$ according to
(\ref{parm}) the canonical choice is
\be
\vece_1 \ = \ \pmatrix{ \sin \beta \cr 
- \cos \beta \cr 0 } \ \ \ \ \ \ \ \
\vece_2 \ = \ \pmatrix{ \cos \beta \cos \gamma \cr
\sin \beta \cos \gamma \cr - \sin \gamma }
\la{epm}
\ee
These vectors
describe a (small) neighborhood ${\cal M} $ 
around $\vecm$ in $\mats^{2}_{\bf m}$. The pre-image 
of ${\cal M}$ under $\vecm(\vecx)$ defines a (thin) 
tubular neighborhood ${\cal T}$ around 
${\cal S}$ in $\matr^3$. At each point 
$\vecx(s)$ along ${\cal S}$ we intersect
the tube ${\cal T}$ by a disk-like surface 
${\cal D}(s)$, normally to each 
of the strings that lie inside ${\cal T}$. For each $s$
the surface ${\cal D}(s)$ then provides a cross-section
of the tube ${\cal T}$. This is mapped onto ${\cal M}$ 
by the vector field $\vecm(\vecx)$. In the limit 
of a very narrow tube ${\cal T}$ the discs ${\cal D}(s)$ 
become surfaces with an area element which is given by 
the pull-back of (\ref{Fij2}), and a unit normal 
which coincides with the pertinent vector $\vect$.

In general the ${\cal D}(s)$ are curved 
in $\matr^3$ and their curvature relates to 
the bending and twisting of the tube ${\cal T}$
in $\matr^3$. A natural local measure of the 
curvature of the surface ${\cal D}(s)$ is given by 
the pull-back of the vector-valued one-form
\be
d \vecm \ = \ - \partial_a
\vecx {B^a}_b \ d u^b \ 
\la{weing}
\ee
where ${B^a}_b$ are the components of
the curvature tensor on $\mats^2_{\bf m} \in 
\matr^3_{\bf m}$ and $u^a$ are local coordinates
in ${\cal M}$. According to (\ref{inde1}),
the components of (\ref{weing}) are indeed
tangential to ${\cal D}(s)$. 
The projections of (\ref{weing}) along the tangent 
vectors of $\mats^2_{\bf m}$
\be
\Phi^{\pm}_i \ = \ \frac{1}{2} \vece_{\pm} \cdot \partial_i \vecm 
\ = \ \frac{1}{2} ( \pm i \partial_i \gamma \ - \ \sin \gamma 
\partial_i \beta)
\la{curv}
\ee
then characterise the local curvature 
of ${\cal D}(s)$ along the directions in $\matr^3$
that are determined by the pre-images of $\vece_{\pm}$.
We propose that (\ref{curv}) is a natural starting point 
for generalizing the curvature (\ref{kappa}) to 
the bundle.

Besides (\ref{u1a}) we now have a second
U(1) gauge transformation. It
acts in the internal $\matr^3_{\bf m}$ by rotating
the basis $\vece_\pm$ according to
\be
\vece_\pm \ \to \ e^{\pm i\chi} \vece_\pm
\la{rotab}
\ee
Since any physical property of the bundle 
should be independent of a particular choice 
of basis vectors (\ref{epm}), the physical
properties of the bundle should also reflect 
gauge invariance under (\ref{rotab}).

The components in (\ref{curv}) are independent
of the coordinate $\alpha$, and hence
manifestly invariant under the U(1) gauge 
transformations (\ref{u1a}) {\it i.e.} 
reparametrizations.
However, these components do not remain
intact under the internal
U(1) gauge transformations (\ref{rotab}). Instead they
transform according to
\[
\vece_{\pm} \cdot 
\partial_i \vecm \ \to \ e^{\pm i \chi} 
\vece_{\pm} \cdot \partial_i \vecm
\]
But there is clearly an intimate relationship
between rotating the points around each other
in ${\cal M}$ and rotating
the strings around each other in ${\cal T}$.
This leads to a relation 
between the configuration space 
U(1) gauge transformations (\ref{u1a}) {\it i.e.} 
reparametrizations of the strings,
and the internal U(1) gauge transformations 
(\ref{rotab}) {\it i.e.} rotations of the
internal space basis vector. Consequently
we redefine (\ref{curv}) into
\be
\Phi^{\pm}_i \ \to \ \kappa^{\pm}_i \ = \
\frac{1}{2} e^{\pm i \alpha} \vece_{\pm} \cdot 
\partial_i \vecm \
\la{curvb}
\ee
These $\kappa^{\pm}_i $ have a non-trivial
weight under both U(1) gauge transformations. But they
remain intact under the diagonal U(1)$\times$U(1) 
gauge transformation where we compensate the internal
U(1) rotation (\ref{rotab}) 
by the configuration space reparametrization 
\be
\alpha \ \to \ \alpha \ - \ \chi
\la{compe}
\ee
We propose that (\ref{curvb}) are the appropriate 
quantities that characterize the curvature 
of the bundle, for the purpose of constructing
energy functionals.

In order to describe the torsion along the bundle
we observe that besides (\ref{curv}), there is one 
additional bilinear that can be constructed from the 
three unit vectors at our disposal
\be
C_i \ = \ 
\frac{i}{2}\vece_{-} \cdot \partial_i \vece_+
\ = \ \cos \gamma \partial_i \beta
\la{b31}
\ee
A comparison with (\ref{tau}) 
suggests that $C_i$ should indeed 
relate to the torsion
along the bundle. However, we find that 
under the frame rotation (\ref{rotab}) the $C_i$
do not remain intact but transform like a U(1) 
gauge field
\[
C_i \ = \ \frac{i}{2}\vece_{-} \cdot \partial_i \vece_+
\ \to \ 
\frac{i}{2}\vece_{-} \cdot ( \partial_i + i \partial_i
\chi ) \vece_+
\ = \ C_i \ - \ \partial_i \chi
\]
Consequently we need to improve (\ref{b31}):
Since the coordinate
$\alpha$ relates to the parameters $s$ along the
individual strings, it becomes
natural to employ the flat connection 
$d\alpha$ in (\ref{Ai2}) to generalize  
(\ref{b31}) into
\[
C_i \ \to \ \tau_i \ = \ \frac{i}{2} \vece_- \cdot 
(\partial_i + i \partial_i \alpha) 
\vece_+ \ = \ \cos\gamma \partial_i
\beta \ - \ \partial_i \alpha 
\]
This should be compared with (\ref{Ai2}).
The U(1) gauge transformation (\ref{rotab})
now acts as follows,
\[
\tau_i \ = \
\frac{i}{2}\vece_- \cdot (\partial_i + i \partial_i \alpha ) 
\vece_+ \ \to \ \frac{i}{2}\vece_- 
\cdot (\partial_i + i 
\partial_i [\alpha + \chi ]) \vece_+
\]
and $\tau_i$ remains invariant
under the diagonal U(1)$\times$U(1)
gauge transformation where we compensate
(\ref{rotab}) by the U(1) gauge transformation 
(\ref{u1a}) that reparametrizes 
the bundle according to (\ref{compe}), 
much like in the case of (\ref{curvb}).
We propose that the $\tau_i$ are
quantities that relate to 
the torsion of the bundle, in a natural
generalization of (\ref{tau}) for the purpose
of constructing energy functionals.

We note that the one-form with components $\tau_i$ has also been 
employed in the differential geometry of surfaces. 
There it is called the connection form \cite{chern} 
and can be shown to be a {\it globally}
defined one-form on the unit tangent bundle 
of $\mats^2$.

We also note that the components in
$\kappa^\pm_i$ and $\tau_i$ are not independent
but there are the following flatness 
relations between the ensuing one-forms
\[
d\tau \ - \ 2i \kappa^+ \wedge \kappa^- \ = \ 
d \kappa^\pm \ - \ \tau \wedge \kappa^\pm \ = \ 0
\]
These reduce the number of independent
field degrees of freedom in $\kappa^\pm_i$, 
$\tau_i$ into three corresponding to the 
torsion, curvature and position along each of
the individual strings.

We now proceed 
to construct an energy functional for the bundle. 
Our starting point is (\ref{kirc}) which suggests
the gauge invariant 
\be
E \ = \ \mu^2 \! \int d^3x \ \left( \
<\!{\kappa}^+ \!, \! {\kappa}^- \!\!> \ + \
c \, | \tau |^2 \ \right) 
\la{kirc2}
\ee
where $\mu$ is a mass scale and $c$ is a dimensionless
constant. Explicitely, the curvature contribution is
\[
<\!{\kappa}^+ \!,\!{\kappa}^- \!\!> \ \equiv \ 
\kappa^+_i \kappa^-_i \ = \
\frac{1}{4}  (\vece_{+} \cdot \partial_i \vecm ) (
\vece_{-} \cdot \partial_i \vecm ) \ = 
\ G^{bc} {B_{ab}}{B^a}_c \ \equiv  \ 
\partial_i \vecm \cdot \partial_i \vecm
\]
where we employ the metric 
$G^{ab} =  \partial_i u^a \partial_i u^b$.
This is manifestly 
invariant under the U(1) gauge transformations. 
The torsion contribution is
\[
 |\tau|^2 \ = \ \frac{1}{4} |\vece_- \cdot 
(\partial_i + i \partial_i \alpha) 
\vece_+ |^2 \ = \ ( \cos\gamma \partial_i \beta 
- \partial_i \alpha)^2
\]
This is also gauge invariant when we compensate
an internal U(1) rotation (\ref{rotab})
by an appropriate reparametrization of the bundle, {\it i.e.} 
by the U(1) gauge transformation (\ref{u1a}).

A scaling argument reveals that (\ref{kirc2}) can
not support bundled strings as knotted solitons.
Consequently we generalize (\ref{kirc2}) into  
\be
E \ = \ \mu^2 \int d^3x \left( \
<\!{\kappa}^+ \!, \!{\kappa}^-\!\!> \ + \
c \, |\tau|^2 \ + \ \lambda_\kappa
<\! d \kappa^+ \!, \!d \kappa^-\!\!> \ + \ 
\lambda_\tau |d \tau|^2 \ \right)
\la{fin1}
\ee
This energy functional has a number of 
interesting properties. For example,
if we remove the terms with explicit dependence 
on the flat connection $d\alpha$ by 
setting $c = \lambda_\kappa = 0$ we arrive 
at
\be
E \ \to \ \mu^2 \! \int d^3x \left(
\ \partial_i \vecm \cdot \partial_i \vecm
\ + \ \frac{\lambda_\tau}{4} 
(\vecm \cdot d\vecm \wedge d \vecm)^2 \ \right)
\la{fadact}
\ee
This is the energy functional of the Faddeev model
\cite{fad1} which is known to 
support stable knots as 
solitons \cite{nature}-\cite{jarmo}. These solitons then
describe stable knotted bundles of closed strings
${\cal S}$.

On the other hand, if we set $c=1$ and $\lambda_\kappa = 
\lambda_\tau$ we arrive at the Hamiltonian of the
Skyrme model which also
supports stable solitons. From the present point
of view the solitons \cite{batsky} of the Skyrme model
are then {\it degenerate} knotted configurations in $\vecm$. 

Both (\ref{fadact}) and the Skyrme Hamiltonian
lead to a relativistically covariant dynamics.
The relation with the Skyrme model also 
tells that we can introduce statistics 
using the Wess-Zumino term \cite{wit}. This yields
the important conclusion that the knotted solitons 
can be quantized either as bosons or as fermions.

For generic values of the parameters we expect 
the solitons of (\ref{fin1}) to be knotted 
and qualitatively quite similar to those discussed in 
\cite{nature}-\cite{jarmo}, except for degenerate 
cases such as the Skyrme model \cite{batsky}.

We note that the functionals $\kappa^\pm_i$
and $\tau_i$ are naturally present in 
a variety of other field theory models such as in the SU(2) 
Yang-Mills theory \cite{omaym} and in a
nonrelativistic model of a two-component plasma \cite{plasma}.
Consequently these theories also admit an
interpretation as a field theory realization  
of bundled strings.  

Finally, we show how we can 
employ field theories
to derive generalizations of the
energy functional (\ref{kirc}).
For this we reduce the energy functional (\ref{fin1}) 
to that of an individual string in the bundle,
by computing the leading contribution to 
the energy of the center-line in the model (\ref{fadact}).

We describe an arbitrary point $\veccx$ in the bundle by
\[
\veccx \ = \ \vecx(s) + r (\cos \theta \vecn + \sin \theta
\vecb) \ \equiv \ \vecx(s) + r\vecu
\]
where $\vecx(s)$ is a coordinate for the center line and
$(r,\theta)$ are plane polar coordinates in the plane 
through $\vecx(s)$, defined by the vectors 
$\vecn$ and $\vecb$. The angle $\theta$ is measured from the
direction of $\vecn$ and $s$ is the arc-length 
measured from an arbitrary point on the center line. When we
move along $\vecx(s)$ by varying the 
parameter $s$, the vector $\vecu$ rotates around the center 
line $\vecx(s)$. We define ${\cal N}$ to be the 
Gauss linking number between the center line
and a generic 
nearby curve of the form $\vecx(s) + \epsilon \vecu(s)$
with some (small) constant $\epsilon$. 
We define $g(s)$ to be an arbitrary function but
with $g(L) = g(0) + L$, and introduce the angle variable
\[
\phi \ = \ \theta + \frac{2 \pi {\cal N} }{L} g(s)
\]
The mapping $s \to g(s)$ implements a reparametrization,
while the use of $\phi$ as an independent variable 
instead of $\theta$ leads to a zero-framing of 
the curve $\vecx(s)$ \cite{moff}. In the coordinates $(r,\phi,s)$ 
any point in a (thin) circular tube ${\cal T}$ with a
(small) radius $R$ surrounding the centerline
can be described by 
\[
\veccx \ = \ \vecx(s) + r \cos[\phi - \frac{2\pi{\cal N}}{L}
g(s) ]  \vecn(s) + r \sin [\phi - \frac{2\pi{\cal N}}{L}
g(s) ] \vecb(s)
\]
The components of the ensuing metric tensor are
\[
d\veccx \cdot d\veccx \ = \ dr^2 + [ 1 - r \kappa \cos (\phi - 
\frac{2\pi{\cal N}}{L} g(s) + r^2 \hat \tau^2 ] ds^2 
+ r^2 d\phi^2 + 2r^2 \hat \tau d\phi ds
\]
where
\[
\hat \tau (s) \ = \ \tau(s) - {g}'(s)
\]
We solve the equations of motion 
that follow from (\ref{fadact}) for
$\beta$ and $\gamma$ in (\ref{parm}) to the leading order $r$ in
the thin circular tube $R\to 0$ limit, and substitute the result to the 
action (\ref{fadact}). When we select the function $g(s)$ so
that it cancels an $s$-dependent contribution to $\gamma$,
we find a static energy functional that involves the following 
universal terms,
\be
E \ = \ a R^2 \int ds | \partial_s
\vecx |^2 \ + \ b R^4 \int ds ( 3 \tau^2(s) + 
R^2 \cdot \kappa^2 \tau^2 ) \ + \ ...
\la{stringe}
\ee
Observe that the energy does not contain a term which involves
the curvature $\kappa$ only, such as the middle term in (\ref{kirc}).
This is entirely consistent with the result that a stable
knotted soliton must have a nontrivial self-linking number:
A curvature contribution such as the middle
term in (\ref{kirc}) scales inversely in the length $L$ of 
the string. Since the first term in (\ref{stringe}) 
is proportional to $L$ the presence of such a 
curvature contribution would lead to a stable circular planar 
configuration with no self-linking.

\vskip 0.5cm

In conclusion, we have studied relations between field
theories and string theories. In particular, we have 
proposed that certain field theories can 
be viewed as effective theories of bundled strings,
which then correspond to knotted solitons in the 
field theory model. We have constructed 
a general class of such field theory models by
employing the natural geometric concepts of torsion
and curvature of curves in $\matr^3$.
In the limit of thin bundles we recover
the usual closed string action with additional torsion
and curvature contributions. Since a field theory
admits a natural particle interpretation, this leads to a
curious dual picture between stringlike excitations and pointlike
particles. We hope that our results lead to a better
understanding of such duality relations between strings and
particles. Indeed, the investigation of these structures
in the context of a Yang-Mills theory might provide
new insight to the properties of colored flux tubes, and 
the appearance of a mass gap and color confinement in the 
theory.

\vskip 1.0cm

We are indebted to Ludvig Faddeev for valuable discussions
and advice, and we also thank A. Alekseev, 
N. Manton and J. Minahan for discussions.

\vfill\eject

\end{document}